# Mössbauer Spectroscopic and XRD Studies of two η-FeAl Intermetallics

Stanisław. M. Dubiel[1a*], Łukasz Gondek[1a], Tilo Zienert[2], J. Żukrowski[1b]

[1]AGH University of Science and Technology, [a]Faculty of Physics and Applied Computer Science, [b]Academic Center for Materials and Nanotechnology, PL-30-059 Kraków, Poland, [2]TU Bergakademie Freiberg, Institute of Material Science (now at Institute of Ceramics, Refractories and Composite Materials), D-09599 Freiberg, Germany

## Abstract

Two intermetallic FeAl compounds with Al content of 70.68 and 72.17 at% were studied using Mössbauer spectroscopy (5–296 K) and X-ray diffraction (15–300 K). The compounds were found to crystallize in the orthorhombic *Cmcm* space group (η-phase). The collected data revealed that dynamics of the Fe atoms (harmonic in entire temperature range) is significantly different that Al atoms. For the latter strong anharmonicity was evidenced. Moreover, it was found that partial filling of the different Al sites leads to occurrence of low- and high- symmetry coordination of Fe atoms, which was reflected in occurrence of two distinct doublets in Mössbauer spectra. All spectral parameters of the doublets as well as the Debye temperature, force constant, kinetic and potential energies of vibrations were determined. Those results revealed significant differences between both alloys, likely originating from approaching the stability boundary of the η-phase for Fe-Al 72.17 at% alloy.

* Corresponding author: Stanislaw.Dubiel@fis.agh.edu.pl

# 1. Introduction

Fe-Al alloys have been of interest both for scientific and industrial purposes. The former follows from the fact that in the system, in addition to a disordered phase ($\alpha$), numerous ordered phases can be obtained viz. $Fe_3Al$(DO$_3$), FeAl(B2), $FeAl_2$($\zeta$), $Fe_2Al_5$($\eta$), $Fe_3Al_7$($\eta''$), $Fe_3Al_8$($\eta'$) and $Fe_4Al_{13}$($\theta$) [1-3]. Each particular compound has its own crystal structure e. g. the orthorhombic $\eta$ (space group *Cmcm*) [4], the monoclinic $\eta'$ (space group *C2/c*) [2] or the orthorhombic long period superlattice structure $\eta''$ (space group *Pmcn*) [3]. Physical, chemical and mechanical properties are characteristic of a given compound. Furthermore, as these compounds can occur within a certain range of composition, their properties can be tuned by changing the composition.

This above-mentioned variety of intermetallics and their structures gives a unique opportunity for experimentalists to study their properties with various methods and techniques. For theoreticians these aluminides are usable to validate their models and calculations e. g. $Fe_2Al_5$ and $Fe_4Al_{13}$ as quasicrystal approximants [5,6].

Regarding the industrial aspect of interest in Fe-Al alloys and compounds one has to mention already used applications like production of aluminized steels. The steels find wide utilization in motor vehicle exhaust systems, domestic appliances and building cladding panels due to their high corrosion resistance combined with a high surface reflectivity [7,8]. The coating layer usually consists of an Al solid solution and/or several intermetallic compounds including $\eta$-$Fe_2Al_5$ or $\theta$-$Fe_4Al_{13}$ [7,9].

Uncontrolled precipitation of the Fe-Al intermetallics in materials can also have a deteriorating effect on their useful properties, so a good knowledge of conditions under which particular compounds can be formed is essential.

The present study was focused on investigation of two FeAl compounds with a composition lying within the phase field of $\eta$, $\eta'$ and $\eta''$ structures [2,3]. The compounds were investigated using Mössbauer spectroscopy that is capable of giving information on electronic properties (charge-densities and electric field gradients experienced by Fe atoms) as well as on lattice dynamical properties (Debye temperature, force constant, kinetic and potential energies of Fe atoms vibrations). To our best knowledge, no results on similar studies were reported in the available literature. Merely, the Debye temperature can be compared with the one determined for η based on low temperature heat capacity measurements [6,10] as well as with that obtained using the first-principles calculations [11].

## 2. Experimental

### 2. 1. Samples

Samples with two nominal compositions 70.68 at.% Al (called thereafter FeAl(I)) and 72.17 at.% Al (called thereafter FeAl(I)) within the homogeneity range of the η-FeAl (~70-73 at.% Al) were prepared by arc melting Al (99.9999) and Fe (99.995). The alloys were then annealed under protective atmosphere of Ar at 1023 K for 42 h and next water quenched. X-ray patterns recorded on powdered samples using a Panalytical Empyrean powder diffractometer with Cu K$_\alpha$ radiation were successfully indexed using the orthorhombic $Fe_2Al_5$ structure reported by Burkhardt et al. [2]. Oxford Cryosystem PheniX closed-cycle cryostat was used for low temperature XRD measurements (20-300K).

### 2. 2. Measurements

$^{57}$Fe Mössbauer spectra were measured in a transmission mode using a standard spectrometer (Wissel GmbH) and a drive operating in a sinusoidal mode. The 14.4 keV gamma rays were supplied by a $^{57}$Co/Rh source. Its activity enabled recording a spectrum of a very good statistical quality within a 1-2 days run. The spectra were recorded in a 1024-channel analyzer in the temperature interval of 5-296K subdivided into two ranges: (a) 5-100K and (b) 80-296K. A closed-cycle Janis Research 850-5 Mössbauer Refrigerator System was used for recording the spectra in the (a)-range whereas for the (b)-range a standard Janis SVT-200 cryostat was employed. The temperature in both cases was kept constant to the accuracy better than ±0.1K. Examples of the measured spectra can be seen in Fig. 1.

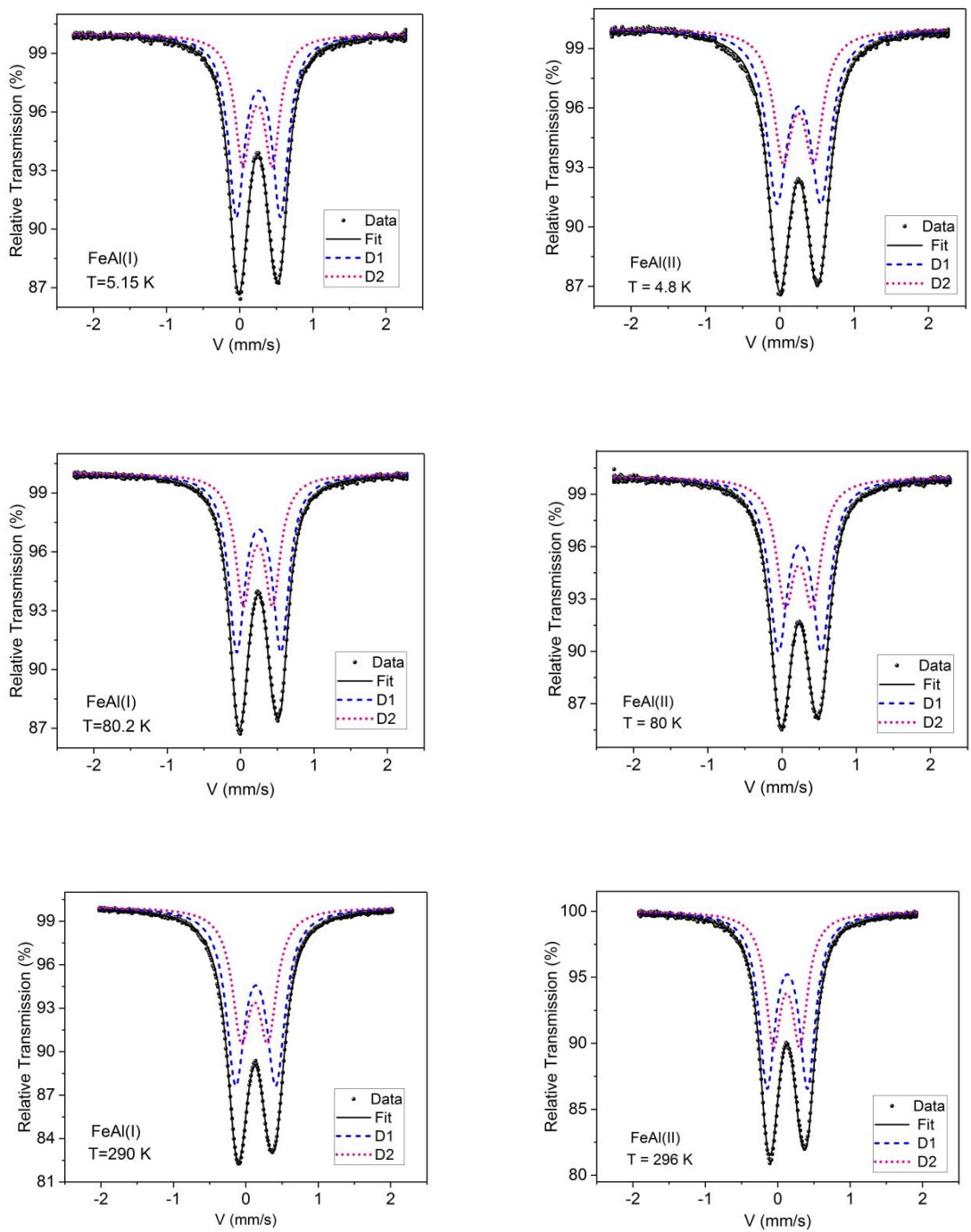

Fig. 1 Examples of the spectra recorded at various temperatures for the studied samples: FeAl(I) – left panel, and FeAl(II) – right panel. The two sub-spectra (doublets) D1 and D2 are marked by dashed and dotted lines, respectively.

## 3. Results and discussion

## 3.1. Spectra analysis

All the measured spectra in form of an asymmetric doublet were analyzed using a least-squares fitting procedure based on a transmission integral method. The asymmetry of the spectra could have been well accounted for by analyzing them in terms of two doublets, D1 and D2, having different center shift, CS1 and CS2, quadrupole splitting, QS1 and QS2, effective thickness, $f'$ and a common linewidth, G. The spectra recorded on both samples were successfully fitted to D1 and D2 assuming 3:2 ratio for their relative abundance (The average ratio obtained treating spectral area of the two doublets as free parameters was 60.3±3:39.7±3). The best-fit values of the spectral parameters obtained with the applied procedure are displayed in Table 1.

Table 1

Best-fit spectral parameters for the studied samples. The meaning of the symbols is given in the text. Temperature of measurements performed in the Janis SVT-200 cryostat Cryostat is marked with a star. Typical errors of the parameters are as follows: $\Delta$CS1=±0.0005 mm/s; $\Delta$CS2=±0.0010 mm/s; $\Delta$QS1=±0.0008 mm/s; $\Delta$QS2=±0.0015 mm/s; $\Delta$G=±0.01 mm/s. Values of the center shift are given relative to the Co/Rh source at room temperature.

| FeAl(I) | | | | | | FeAl(II) | | | | | |
|---|---|---|---|---|---|---|---|---|---|---|---|
| T/K | CS1 | CS2 | QS1/2 | QS2/2 | G | T/K | CS1 | CS2 | QS1/2 | QS2/2 | G |
| 5.15 | .2556 | .2404 | .2887 | .1912 | 0.25 | 4.8 | .2480 | .2343 | .2852 | .1868 | 0.26 |
| 11.3 | .2557 | .2387 | .2887 | .1914 | 0.25 | 11.1 | .2475 | .2340 | .2850 | .1869 | 0.25 |
| 15.8 | .2552 | .2398 | .2885 | .1913 | 0.25 | 15.2 | .2478 | .2331 | .2851 | .1868 | 0.25 |
| 25.4 | .2543 | .2395 | .2884 | .1911 | 0.25 | 20.4 | .2478 | .2334 | .2849 | .1866 | 0.25 |
| 30.3 | .2541 | .2396 | .2879 | .1912 | 0.25 | 30.0 | .2477 | .2336 | .2848 | .1867 | 0.25 |
| 40.2 | .2538 | .2390 | .2881 | .1909 | 0.25 | 40.2 | .2467 | .2313 | .2849 | .1866 | 0.25 |
| 50.2 | .2536 | .2387 | .2880 | .1908 | 0.25 | 50.2 | .2467 | .2328 | .2846 | .1865 | 0.25 |
| 60.3 | .2525 | .2377 | .2877 | .1907 | 0.25 | 60.2 | .2459 | .2307 | .2844 | .1863 | 0.25 |
| 70.0 | .2501 | .2354 | .2874 | .1902 | 0.25 | 70.0 | .2434 | .2281 | .2842 | .1862 | 0.25 |
| 80.2 | .2476 | .2329 | .2871 | .1900 | 0.25 | 80.0 | .2409 | .2244 | .2840 | .1860 | 0.25 |
| 80* | .2477 | .2328 | .2869 | .1902 | 0.24 | 80.0* | .2400 | .2266 | .2839 | .1857 | 0.24 |
| 90.2 | .2457 | .2311 | .2866 | .1905 | 0.25 | 90.1 | .2388 | .2232 | .2837 | .1856 | 0.25 |
| 95* | .2439 | .2282 | .2865 | .1900 | 0.23 | 95* | .2365 | .2247 | .2835 | .1854 | 0.24 |
| 100.2 | .2433 | .2278 | .2863 | .1897 | 0.25 | 100.1 | .2361 | .2219 | .2834 | .1852 | 0.25 |
| 110* | .2370 | .2208 | .2858 | .1893 | 0.23 | 110* | .2317 | .2190 | .2832 | .1850 | 0.23 |
| 125* | .2317 | .2156 | .2854 | .1890 | 0.23 | 125* | .2266 | .2142 | .2830 | .1846 | 0.23 |

| 140* | .2255 | .2087 | .2849 | .1885 | 0.23 | 140* | .2207 | .2075 | .2829 | .1842 | 0.23 |
| 155* | .2187 | .2015 | .2841 | .1880 | 0.23 | 155* | .2138 | .2014 | .2818 | .1939 | 0.23 |
| 170* | .2111 | .1950 | .2829 | .1875 | 0.23 | 170* | .2072 | .1941 | 0.2814 | 0.1837 | 0.23 |
| 185* | .2044 | .1855 | .2826 | .1867 | 0.23 | 185* | .1987 | .1873 | 0.2803 | 0.1833 | 0.23 |
| 200* | .1967 | .1773 | .2812 | .1859 | 0.24 | 200* | .1913 | .1788 | 0.2797 | 0.1828 | 0.23 |
| 215* | .1878 | .1700 | .2800 | .1852 | 0.24 | 215* | .1836 | .1709 | 0.2786 | 0.1822 | 0.24 |
| 230* | .1799 | .1612 | .2795 | .1845 | 0.24 | 230* | .1753 | .1625 | 0.2781 | 0.1819 | 0.24 |
| 245* | .1708 | .1535 | .2790 | .1835 | 0.24 | 245* | .1675 | .1541 | 0.2775 | 0.1809 | 0.23 |
| 260* | .1603 | .1465 | .2780 | .1824 | 0.24 | 260* | .1597 | .1457 | 0.2757 | 0.1805 | 0.24 |
| 275* | .1536 | .1354 | .2770 | .1816 | 0.24 | 275* | .1506 | .1381 | 0.2742 | 0.1798 | 0.24 |
| 290* | .1443 | .1231 | .2761 | .1808 | 0.24 | 296* | .1392 | .1275 | 0.2732 | 0.1795 | 0.24 |

### 3.2. Debye temperature

The Debye temperature, $T_D$, is regarded as a fundamental parameter in solid state physics. It is pertinent to various physical properties of materials, such as specific heat, elastic constants and melting point, hence its knowledge is important. It can be experimentally determined using different methods including Mössbauer spectroscopy. With the latter $T_D$ can be figured out either from a temperature dependence of a center shift, CS, or that from a recoil-free fraction, f. The former dependence can be written by the following expression:

$$CS(T) = IS(T) + SOD(T) \qquad (1)$$

Where IS stays for the isomer shift and SOD is the so-called second order Doppler shift i.e. a quantity related to a non-zero mean value of the square velocity of vibrations, $<v^2>$, hence a kinetic energy. Assuming the phonon spectrum obeys the Debye model, and that IS hardly depends on temperature, so it can be ignored [12], the temperature dependence of CS goes practically via the second term which is related to $T_D$ via the following relationship [13]:

$$CS(T) = IS(0) - \frac{3k_B T}{2mc}\left( \frac{3T_D}{8T} + 3\left(\frac{T}{T_D}\right)^3 \int_0^{T_D/T} \frac{x^3}{e^x - 1} dx \right) \qquad (2)$$

Here m stays for the mass of the Fe atom, $k_B$ is the Boltzmann constant, c is the speed of light, and $x = \frac{\hbar\omega}{2\pi k_B T}$ ($\omega$ being frequency of vibrations).

Figure 2 illustrates *CS(T)* dependences found for FeAl(I). Values of the Debye temperature obtained for both samples based on eq. (2) are presented in Table 2.

Table 2.

Values of the Debye temperature, $T_D$, as determined from the temperature dependence of the center shift, *CSk*, of particular components (k=1, 2), and that of the average center shift, <CS>, as well as from the quantity proportional to the recoil-free fraction, *f*.

|  | FeAl(I) | | | |
|---|---|---|---|---|
|  | CS1 | CS2 | <CS> | f'/f$_o$' |
| $T_D$ (K) | 456(5) | 433(5) | 449(5) | 344(4) |
|  | FeAl(II) | | | |
| $T_D$ (K) | 485(6) | 496(5) | 489(5) | 350(4) |

Alternatively, $T_D$ can be calculated from a temperature dependence of the recoil-free fraction, $f=exp(-k^2<x^2>)$, k being the wave vector of the gamma rays and $<x^2>$ stands for the mean square amplitude of vibrations. In the frame of the Debye model the $f$-$T_D$ relationship reads as follows [14]:

$$f = \exp\left[\frac{-6E_R}{k_B T_D}\left\{\frac{1}{4} + \left(\frac{T}{T_D}\right)^2 \int_0^{2T_D/T} \frac{xdx}{e^x - 1}\right\}\right] \qquad (3)$$

Where $E_R$ is the recoil kinetic energy, $k_B$ is Boltzmann constant.

The $f$ – factor is proportional to the so-called effective thickness, $f'$. Values of the latter can be obtained by analyzing spectra in terms of the transmission integral method as was the present case.

The values of $T_D$ achieved in this way are also included in Table 2. Their values are significantly smaller than those found from the $CS(T)$ dependences. This effect follows from the fact that the center shift depends on the mean square velocity of vibrations, hence the kinetic energy, while the recoil-free fraction is related to the mean square amplitude of vibrations, hence the potential energy (in the harmonic motion approximation). The two spectral parameters viz. $CS$ and $f$ are sensitive to different frequencies of vibrations, hence a different parts of the phonon density of states (PDS). The shape of the latter is, in general, not parabolic and consequently, values of $T_D$ derived from $CS$ and $f$ are usually different.

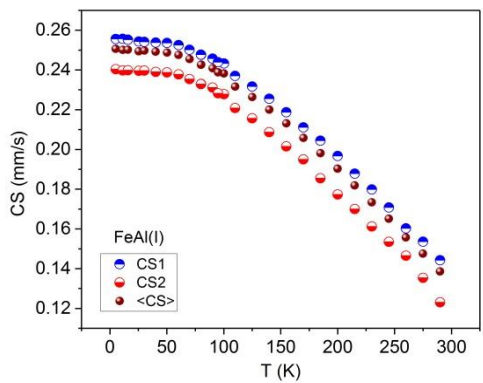 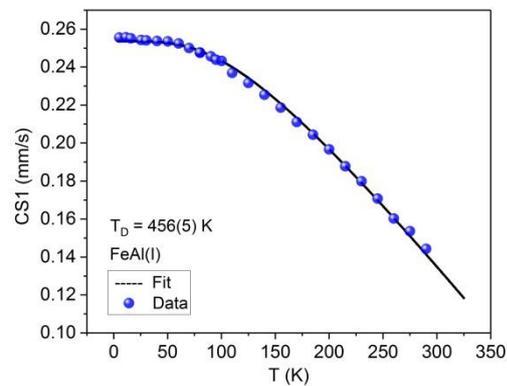

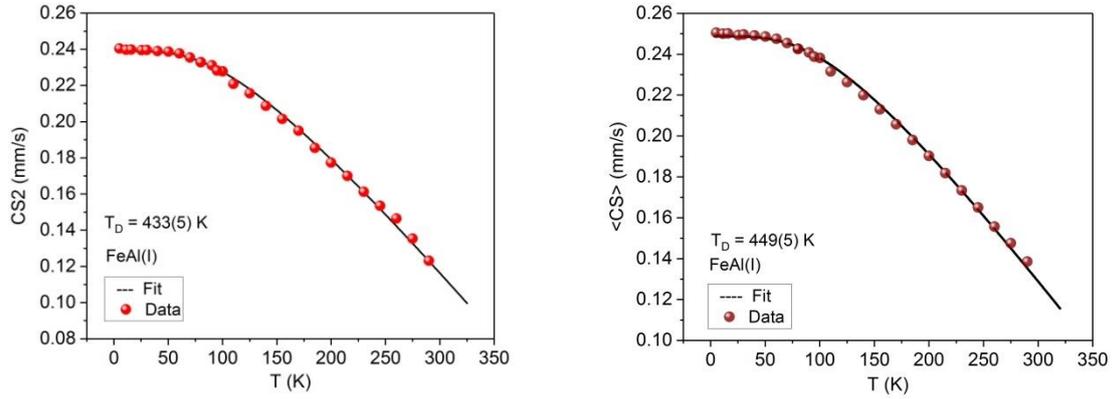

Fig. 2 Temperature dependences of the center shift for the FeAl(I) sample: (upper left) experimental data for CS1, CS2 and <CS>, experimental data for CS1, CS2 and <CS> together with the best-fit of eq. (1) to the data. Values of the Debye temperature, $T_D$, derived therefrom are indicated.

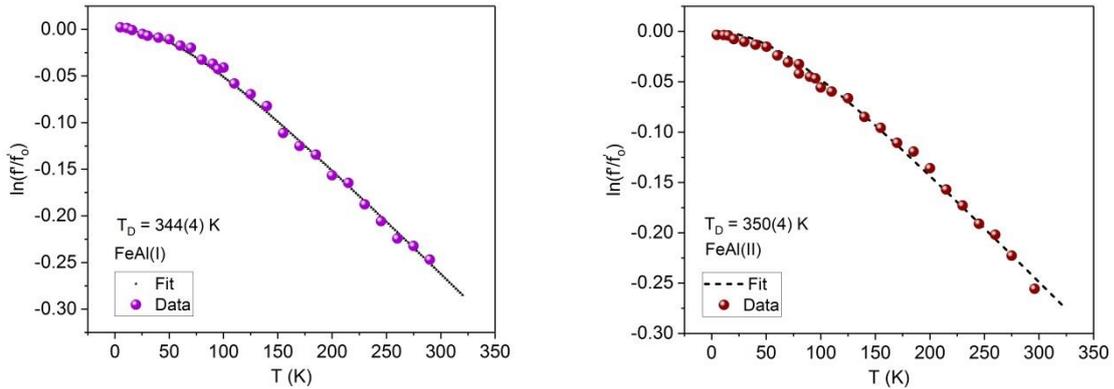

Fig. 3 Temperature dependences of $ln(f'/f_o^{'})$ for the studied samples ($f'_o$ being $f'$ at the lowest temperature). The best-fits of the data to eq. (3) are indicated, too. Values of the Debye temperature, $T_D$, are displayed.

Noteworthy, the values of $T_D$ found from the center shift for the FeAl(II) sample, hence the one with a slightly higher content of Al are by ~10% higher than the corresponding ones revealed for the FeAl(I) sample. It should be also noticed that our values of $T_D$ agree well with the one found from heat capacity measurements for $Fe_2Al_5$ [6,10], but they are significantly smaller than the Debye temperature of ~668 K calculated for the Fe$_2$Al$_5$ intermetallic compound [11]. It should be, however, mentioned that inconsistencies between values of $T_D$ for a given material as determined with different methods are known because, in general, shapes of phonon density of states (PDS) are not parabolic. Various experimental methods are sensitive to different parts of PDS yielding consequently different $T_D$-values.

### 3. 3. Energetics of Vibrations

The vibrations of Fe atoms in the studied samples can be expressed in terms of the underlying kinetic, $E_K$, and potential, $E_P$, energies. The average kinetic, $E_K=0.5m<v^2>$, and potential, $E_P=0.5F<x^2>$ (in harmonic approximation) energies of the lattice vibrations can be determined if the $SOD$ and the $f$-factor are known. The force constant is denoted by $F$, $m$ is the mass of an vibrating atom (here $^{57}$Fe) and $c$ is the velocity of light. Taking into account that by definition $SOD=-0.5E_\gamma<v^2>/c^2$, $E_\gamma$ being the energy of the gamma-rays (14.4 keV in the present case), the average kinetic energy can be expressed as follows:

$$E_K = -mc^2 \frac{SOD}{E_\gamma} \qquad (4)$$

The connection between $E_P$ and $f$ is, in turn, given by the following equation:

$$E_P = -\frac{1}{2} F \left( \frac{\hbar c}{E_\gamma} \right)^2 \ln f \qquad (5)$$

### 3. 3. Energetics of Vibrations

#### 3. 3. 1. Kinetic energy

$E_K$ can be easily calculated based on eq. (4) and using the $SOD$-values measured in the Mössbauer experiment. However, in order to calculate $E_P$ based on eq. (5) one has to know the value of the force constant, *F*. The calculated changes of $E_k$ based on formula (4), $\Delta E_k$, are plotted vs. temperature in Fig. 5.

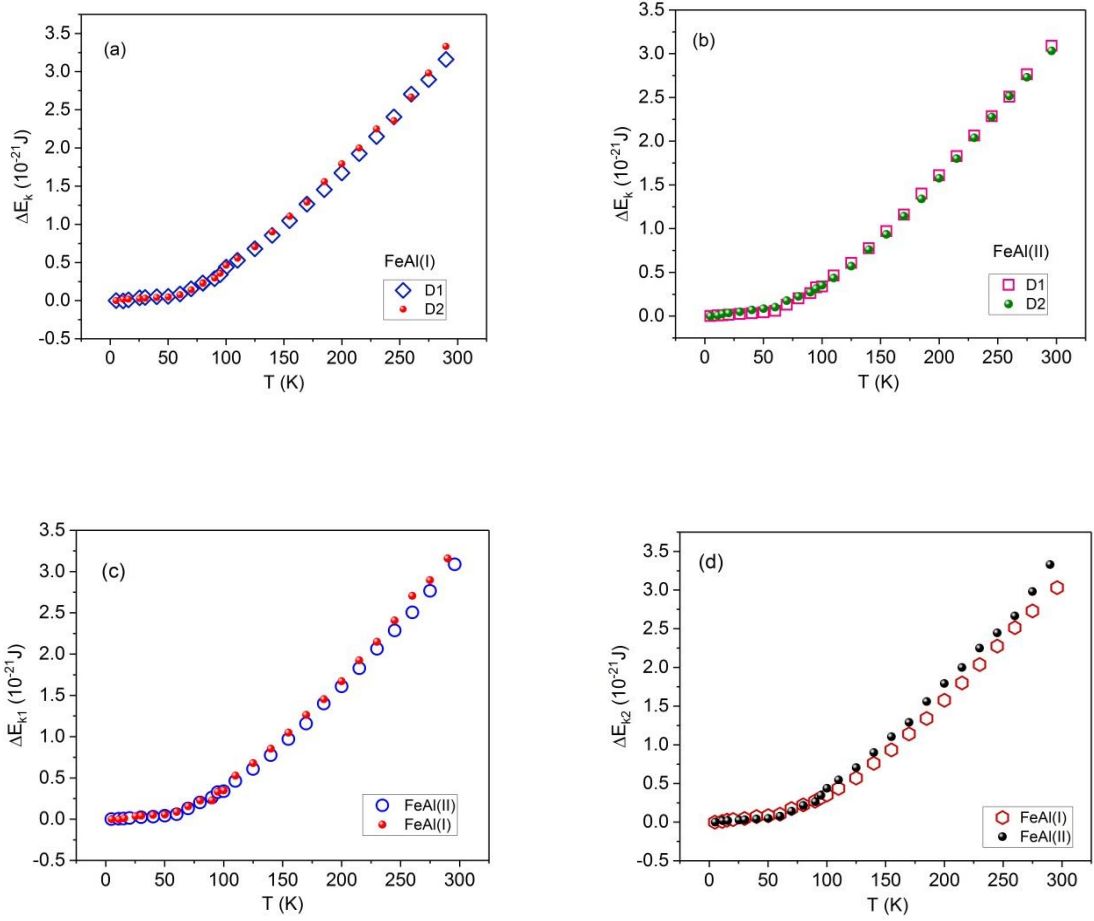

Fig. 5 Change of the kinetic energy, $\Delta E_k = E_k(T) - E_k(5K)$, vs. temperature, *T*, for both lattice sites (represented by doublets D1 and D2) in: (a) FeAl(I), (b) FeAl(II), and the change of the kinetic energy in both samples for the site: (c) D1 (8f), $\Delta E_{k1} = E_{k1}(T) - E_{k1}(5K)$, and (d) D2 (4e), $\Delta E_{k2} = E_{k2}(T) - E_{k2}(5K)$.

It can be concluded from the plots shown in Fig. 5 that the kinetic energy changes (1) for a given sample are the same for both spectral components, hence lattice sites, and (2) exhibit some small differences for T > ~100 K for a given spectral component in different samples.

## 3. 3. 2. Potential energy

As already mentioned the knowledge of the force constant, $F$, is necessary for determining $E_p$ within the harmonic model approximation. Based on the present experiment, $F$ can be figure out if the $\Delta<v^2>$-$\Delta<x^2>$ relationship is linear. In such case $F=m\cdot\alpha$, where $\alpha$ is a slope of the $\Delta<v^2>$-$\Delta<x^2>$ relationship. It has turned out that in a wide temperature interval viz. for $T>\sim60$ K for all four cases (two sub-spectra for two samples) the relationship is linear. Two pertinent examples are presented in Fig. 6.

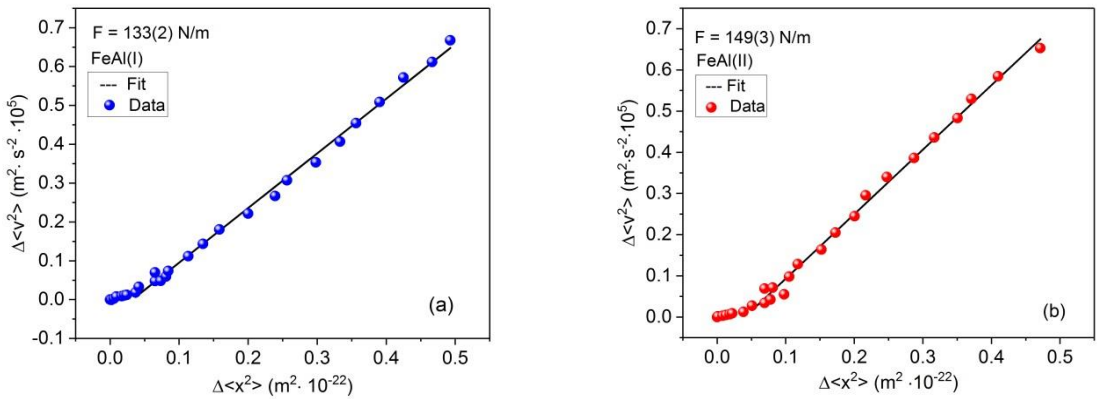

Fig. 6 Relationship between $\Delta<v^2>$ and $\Delta<x^2>$ for the sub-spectrum D1 in: (a) FeAl(I) and (b) FeAl(II). The data in a linear parts were fitted to a linear equation and the best fits are marked by solid lines. Derived therefrom values of the force constant, $F$, are shown in legends.

In this way the following values of the force constant were obtained: $F_1$=133(2) N/m and $F_2$=138(2) N/m for the major (D1) and for the minor (D2) sites in FeAl(I), respectively. For the FeAl(II) compound the corresponding $F$-values are: $F_1$=149(3) N/m and $F_2$=150 N/m. Thus it follows that the more Al-concentrated FeAl intermetallic has by ~10% higher values of the Debye temperature as well as those of the force constant. Noteworthy, the presently found values of the force constant are similar to that determined for a σ-phase FeCr intermetallic compound using the nuclear resonant inelastic x-ray scattering method [15].

In the case of harmonic oscillations changes of the two forms of the mechanical energy i.e. $E_k$ and $E_p$ should be exactly the same within the same temperature range. In order to see whether or not this is the case here, a relationship between $\Delta E_k$ and $\Delta E_p$ for each spectral component (lattice site) was calculated. It has turned out that for each of the two sites in both samples the $\Delta E_k$ - $\Delta E_p$ relationship is linear (except small temperature range < ~50-60 K) and the slope of each line is close to 1. This means that $\Delta E_k = \Delta E_p$ i.e. the lattice vibrations in the studied intermetallics are harmonic. Examples of the $\Delta E_k$ - $\Delta E_p$ correlation can be seen in Fig. 7.

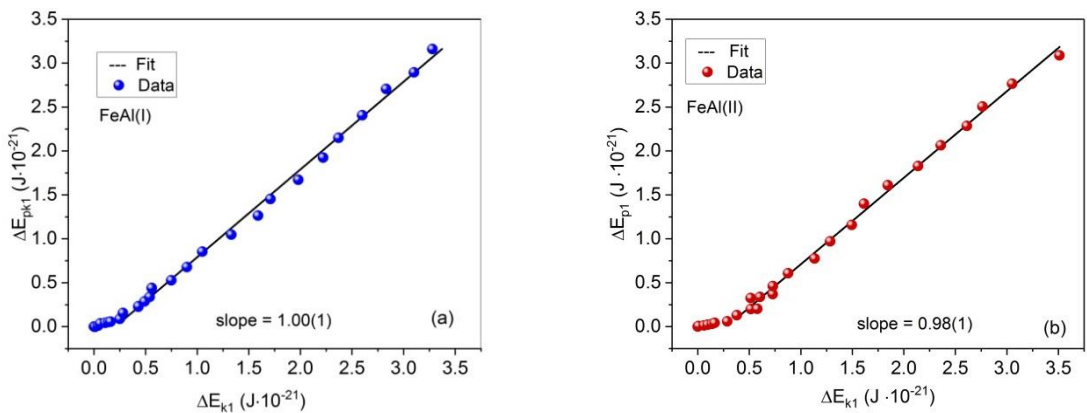

Fig. 7 Examples of a relationship between $\Delta E_k$ and $\Delta E_p$ for the major component (D1) in: (a) FeAl(I), and (b) FeAl(II). The lines represent the best linear fit to the data for $T$>~50-60 K.

### 3.4. Quadrupole splitting

Figure 8 illustrates temperature dependences of QS1, QS2 and <QS> for both samples. It is evident that QS1 is significantly greater than QS2 which means that the electric field gradient sensed by Fe atoms represented by D1 is greater than the one acting on Fe atoms associated with D2.

The temperature dependence of the quadrupole splitting, *QS(T)*, can be well fitted to the following phenomenological equation [16,17]:

$$QS(T) = QS(T_o)\left[1 - aT^{3/2}\right] \quad (4)$$

Figure 8 gives evidence that the temperature dependence of the presently obtained QS-values is in line with this equation, too. The parameter "a" has the value of $1 \cdot E-5$ $K^{-3/2}$ which is typical of metallic systems.

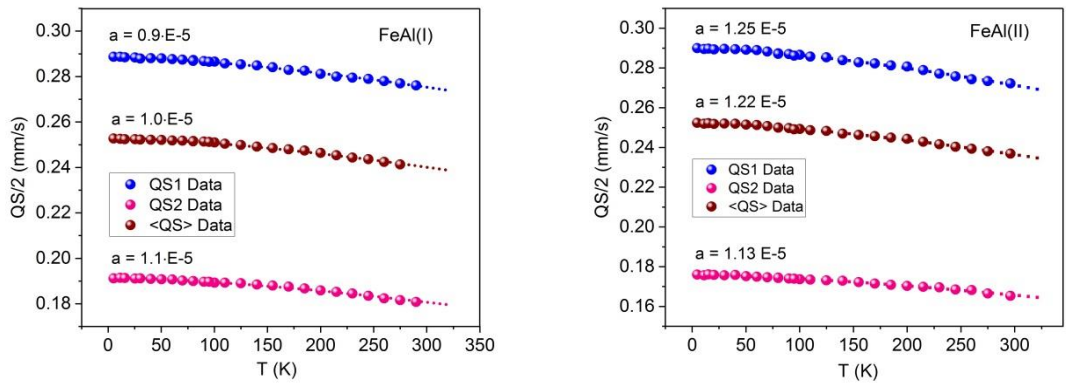

Fig. 8 Temperature dependences of the quadrupole splitting together with the best-fits of the data to eq. (4). Values of the "a" parameter, in $K^{-3/2}$, are shown for each case.

### 3.5. X-ray diffraction

Auxiliary X-ray diffraction (XRD) studies were performed in a wide temperature range to strengthen conclusions drawn from Mössbauer spectroscopy. In Figure 10 XRD exemplary pattern of FeAl(II) and the unit cell volume versus temperature are presented. Room temperature patterns revealed that both samples can be indexed within the orthorhombic unit cell of the *Cmcm* space group. Fe atoms occupy the 4c site, while the Al atoms reside on the 8g, 8f and 4b sites. The two latter sites turned out to be only partially filled. No traces of additional reflections originating from the η' phase can be observed. Rietveld method [18] was used to refine lattice parameters of the investigated samples and they are given in Table 3. Apparently, the unit cell volume is larger for the sample with a higher Al content, what seems to be in line with the difference in metallic radiuses of Al (143 pm) and Fe (126 pm). Noteworthy, the values of the lattice parameters are in line with experimental ones given in [11].

Table 3 Lattice parameters and isotropic Debye-Waller factors at 300 K as derived from XRD studies.

|  | a [Å] | b [Å] | c [Å] | Fe $B_{ISO}$ [Å$^2$] | Al $B_{ISO}$ [Å$^2$] |
|---|---|---|---|---|---|
| FeAl(I) | 7.6535(2) | 6.4137(2) | 4.2133(1) | 0.14(1) | 0.76(3) |
| FeAl(II) | 7.6561(2) | 6.4107(2) | 4.2205(1) | 0.13(1) | 0.68(2) |

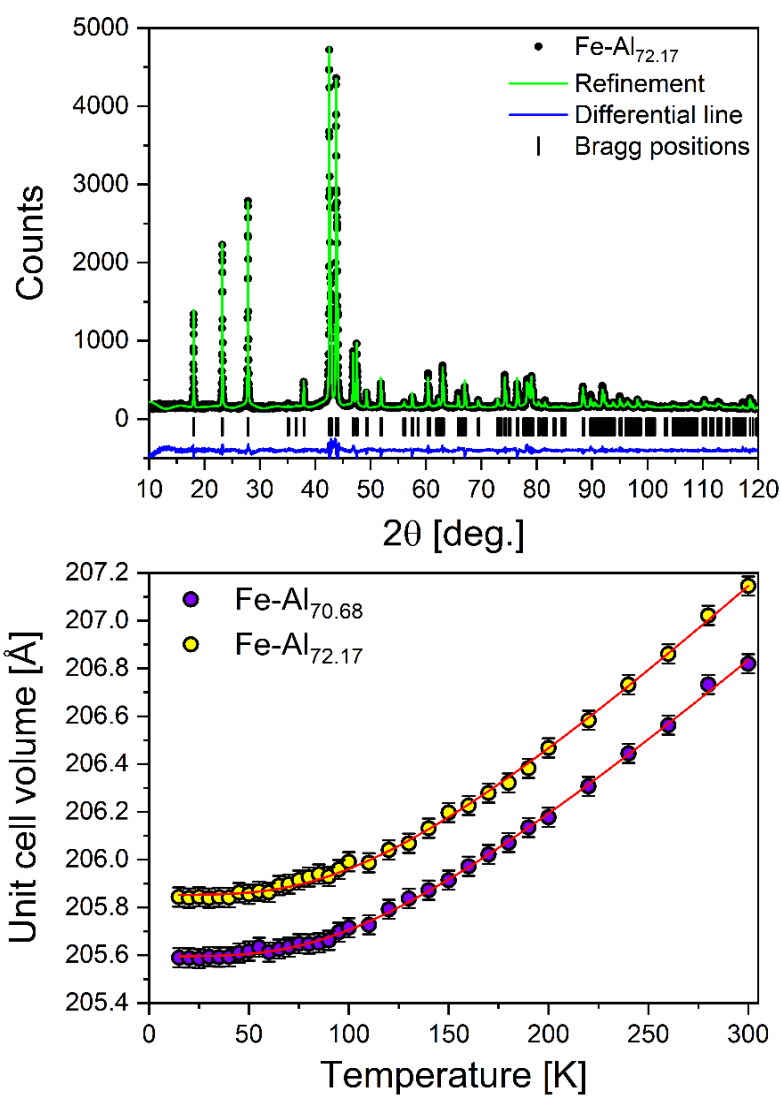

Fig. 10 X-ray diffraction studies of FeAl(I) and FeAl(II). Red lines in the bottom panel mark the fits of the Debye model to the data.

The coordination of Fe atom which entirely occupy the 4c site is quite complex, as presented in Fig. 11. There are three sites available for Al atoms, however only one of them (8g) is fully occupied. The remaining 8f and 4b sites are filled up to 25% and 33%, respectively. As apparent from Fig. 11, the Fe atom coordination may differ due to statistical filling of the 8f and 4b sites. Due to Al atomic size there is no possibility of simultaneous filling of all 8f and 4b sites. Taking this into account, low- and high-symmetry coordination of Fe can be distinguished. Some examples of such coordination are presented in Fig. 11. Rough estimation based on the site occupation probability and possibility of simultaneous filling of adjacent Al sites leads to conclusion that the ratio between the low and the high symmetry coordination is close to 3:2.This is consistent with the occurrence of two different contributions to the Mössbauer spectra consisting of high and low quadrupole splitting components, indicative of dissimilar electric field gradients experienced by Fe probe atoms.

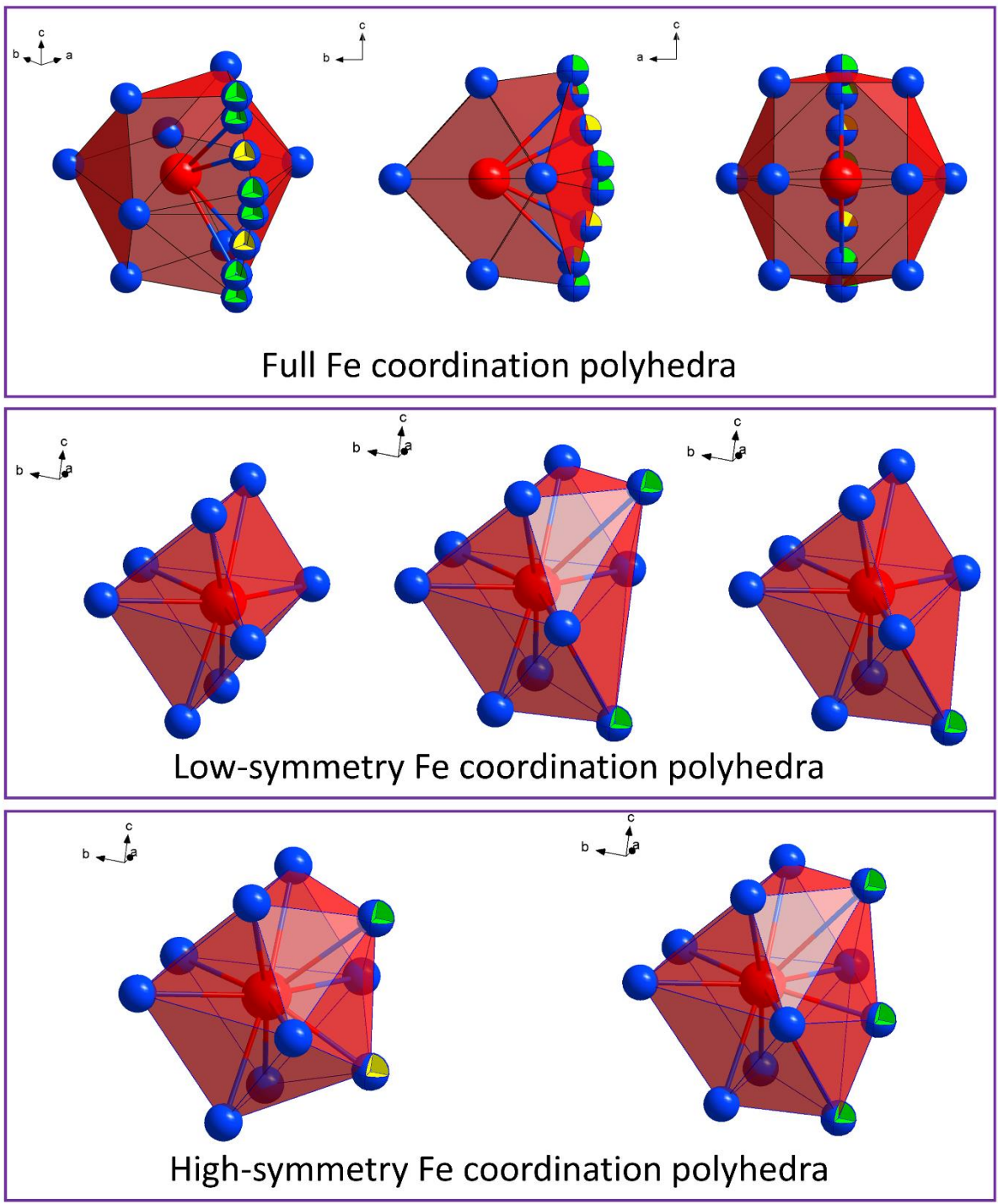

Fig. 11 Coordination of Fe atom at the 4c site (red). Al atoms are marked as: blue (at the 8g site); blue/green (at the 8f site) and blue/yellow (at the 4b site). Examples of a low- and a high-symmetry Fe coordination are presented according to partial filling of the 8f and 4b sites.

At low temperatures only monotonic changes in lattice parameters were observed for both investigated specimens. Unit cell volumes of FeAl(I) and FeAl(II) are shown in Fig. 10 (the bottom diagram). Their temperature dependencies can be approximately described using the Debye formula [19]:

$$V = V_0 + I_C \frac{T^4}{\theta_D^3} \int_0^{\frac{\theta_D}{T}} \frac{x^3}{e^x - 1} dx \quad (5)$$

where: $V_0$ is the unit cell volume at 0 K, $I_C$ is the coefficient including the Grüneisen and compressibility parameters and $\theta_D$ is the Debye temperature. For temperatures above roughly 200 K the $I_C$ coefficient is the slope of $V(T)$ dependence.

Table 4 Results of the refinement of equation (5) to the unit cell volume versus temperature.

|  | $\theta_D$ [K] | $I_C$ [Å³/K] | $V_0$ [Å³] |
| --- | --- | --- | --- |
| FeAl(I) | 449(7) | 0.0226(1) | 205.59(1) |
| FeAl(II) | 471(8) | 0.0243(1) | 205.85(1) |

One can notice marginal lattice expansion below ~50 K (see Fig. 10), which suggest that atomic vibrations are mostly harmonic. At higher temperatures atoms vibrations become anharmonic, resulting in typical thermal expansion of the lattice. In context of linear dependence between kinetic and potential energy for Fe atoms in the entire temperature range evidenced by the Mössbauer spectroscopy (see Fig. 7), it seems that vibrations of lighter Al atoms are mainly responsible for anharmonic effects for the entire crystal. This conclusion is supported by the behavior of isotropic Debye-Waller factors, presented in Table 3. The $B_{ISO}$ factor, proportional to the mean-square amplitude of thermal vibrations, is 5 times higher for Al than for Fe at 300 K.

## 4. Conclusions

The results of XRD and Mössbauer spectroscopy measurements made for two Fe-Al intermetallics with Al content of 70.68 and 72.12 at.% let us to draw following conclusions:

1. Both investigated samples are single-phase with the orthorhombic structure known as η-$Fe_2Al_5$ (space group *Cmcm*).

2. The 4b and 8f positions are only partially occupied by Al atoms, leading to occurrence of low- and high-symmetry Fe coordination polyhedra.

3. Mössbauer spectra are asymmetric and they could be successfully analyzed in terms of two doublets: major D1 (low-symmetry coordination) and minor D2 (high-symmetry coordination).

4. Debye temperature was determined from temperature dependences of: the center shifts associated with D1 and D2; the effective thickness (recoil-free fraction) and the unit cell volume. It was found that the Debye temperatures derived from the center shifts and the XRD measurements are in a good agreement. The Debye temperature rises significantly with the increase of Al content.

5. Force constant values depicting Fe atoms in two different atomic environments (D1 and D2) were estimated.

6. Changes in the kinetic and potential energies of Fe atom vibrations were determined. Vibrations of Fe atoms are harmonic up to the 300 K, while the overall lattice anharmonicity is led by Al atoms.

7. Temperature dependences of the quadrupole splitting were found to be in line with the $T^{3/2}$ law.

**Acknowledgements**

This work was financed by the Faculty of Physics and Applied Computer Science AGH UST and ACMIN AGH UST statutory tasks within subsidy of Ministry of Science and Higher Education, Warszawa and by the German Research Foundation (DFG) in frame of the subproject A03 within the Collaborative Research Centre SFB 920.

**CRediT author statement**

**Stanisław M. Dubiel:** Conceptaulization, Methodology, Validation, Formal Analysis of Mössbauer Spectra, Investigation, Data Curation, Writing - Original Draft, Supervison, Funding acquisition; **Łukasz Gondek:** Measurements of XRD Patterns and their formal Analysis, Validation, Writing - Review & Editing; **Tilo Zienert:** Resources, Writing - Review & Editing; **Jan Żukrowski:** Software, Investigation, Writing - Review & Editing.

**References**

[1] U. Kattner and B. Burton, Phase Diagrams of Binary Iron Alloys, ASM International, 1993, pp. 12-28, Chapter Al-Fe.

[2] N. L. Okamoto, J. Okumura, M. Higashi, H. Inui, Acta Mater., 29 (2017) 290.

[3] N. L. Okamoto, M. Higashi, H. Inui, Sci. Techn. Adv. Mater., 20 (2017) 543.

[4] U. Burkhardt, Yu. Grin, M. Ellner, K. Peters, Acta Crystallogr. Sect. B, 50 (1994) 313.

[5] J. Grin, U. Burkhardt, and M. Ellner, Z. Kristallogr., 209 (1994) 479

[6] J. Chi, X. Zheng, S. Y. Rodriguez, Phys. Rev. B, 82 (2010) 174419

[7] R.W. Richards, R.D. Jones, P.D. Clement et al., Int. Mater. Rev., 39 (1994) 191.

[8] Y. J. Li, J. Wang, H. Q. Wu, Mater. Res. Bull., 36 (2001) 2389.

[9] N. Takata, M. Nishimoto, S. Kobayash et al., Intermetallics, 67 (2015) 1.

[10] T. Zienert, A. Leineweber, O. Fabrichnaya, J. Alloy. Comp., 725 (2017) 848.

[11] Y. H. Liu, X. Y. Chong, Y. H. Jiang et al., Physica B, 506 (2017) 1.

[12] S. Willgeroth, H. Ulrich, J. Hesse, J. Phys. F: Met. Phys. 14 (1984) 387

[13] K. N. Shrivastava, Hyperfine Interact., 26 (1985) 817

[14] N. N. Greenwood, T. C. Gibb, in *Mössbauer Spectroscopy,* Chapman and Hall Ltd, London, 1971

[15] S. M. Dubiel, J. Cieślak, W. Sturhahn et al., Phys. Rev. Lett., 104 (2010) 155503

[16] E. N. Kaufmann, R. H. Vianden, Rev. Mod. Phys., 51 (1979)161

[17] K. Al-Qadi, P. Wang, Z. M. Stadnik, J. Przewoźnik, Phys. Rev. B, 79 (2009) 224202

[18] J. Rodriguez-Carvajal, Fullprof Program, Physica B 192 (1993) 55

[19] F. Sayetat, P. Fertey, M. Kessler, J. App. Cryst. 31 (1998) 121